\documentclass[doublespacing]{elsart}
\usepackage{epsfig}

\begin{document}

\begin{frontmatter}

\title{Multiscale SOC in turbulent convection}
\author[label1]{K.R. Sreenivasan\corauthref{col1}}
\author[label1,label2]{A. Bershadskii}
\author[label1]{J.J. Niemela}
\address[label1]{The Abdus Salam International Centre for Theoretical
Physics, Strada Costiera 11, I-34100 Trieste, Italy}
\address[label2]{ICAR, P.O. Box 31155, Jerusalem 91000, Israel}
\corauth[col1]{Corresponding author: K.R. Sreenivasan, E-mail:
krs@ictp.trieste.it.}

\begin{abstract}
Using data obtained in a laboratory thermal convection experiment
at high Rayleigh numbers, it is shown that the multiscaling
properties of the observed mean wind reversals are quantitatively
consistent with analogous multiscaling properties of the
Bak-Tang-Wiesenfeld prototype model of self-organized criticality
in two dimensions.

\end{abstract}


\end{frontmatter}

\section{Introduction}
\label{1} The existence of well-organized large-scale motion in
turbulent thermal convection is an intriguing recent discovery. It
has been under active investigation for some time, both
theoretically and experimentally (see \cite{kh}-\cite{vc} and
references therein). At least in a convection apparatus whose
aspect ratio (= the ratio of the diameter to the height) is unity,
the large-scale motion is in the form of a persistent circulation
(on the average), its physical extent being of the order of the
size of the apparatus itself. This is often called the ``mean
wind". The mean wind evolves with the Rayleigh number (which can
be regarded as an external tunable parameter in the flow), and
seems to asymptote to a well-defined shape only beyond Rayleigh
numbers of the order of $10^{10}$. However, beginning more or less
at this Rayleigh number, at least for three or so further decades,
the mean wind undergoes abrupt and apparently irregular reversals
of direction. These reversals are the object of the study here.

The measurements to be analyzed were obtained at a Rayleigh number
of $1.5 \times 10^{11}$ in thermal convection occurring in a
closed container of circular cross-section of 50 cm diameter and
aspect ratio unity. The working fluid was cryogenic helium gas.
The same data have been analyzed in the past \cite{sbn}, where the
fluid dynamical origin of reversals was proposed to be the
imbalance between buoyancy and friction (with inertia playing a
secondary role). It was further suggested that a stochastic
``change of stability" occurs between two metastable states
corresponding to the opposite direction of the wind. The main
physical quantity analyzed in \cite{sbn} was the interval $\tau$
(the life-time of the metastable states, or the interval between
reversals). The time of actual switching between the two states
was short on the scale of some average measure of $\tau$, so the
wind reversals could be regarded as abrupt. The last property
(perhaps also the precise details of reversals themselves) is most
likely dependent on the specific boundary conditions of the
experiment, but it is thought that the statistical properties of
the duration times $\tau$ are insensitive to these details. The
analysis of \cite{sbn} indicated a general dynamic mechanism
similar to self-organized criticality (SOC) \cite{bak}-\cite{ms}.
The particular emphasis of this paper is the elaboration of this
analogy.

Self-organized criticality occurs through a nonlinear feedback
mechanism. There could be many possible scenarios of SOC in our
system where numerous plumes and jets are developed as a result of
boundary layer and thermal instabilities, all of which are
embedded in a background of strong turbulent fluctuations
prevalent in the core. It was suggested recently for plasma
turbulence (see Refs. \cite{dh},\cite{cnld}) that instabilities
governed by a threshold may lead to self-organized criticality by
producing transport events at all scales (avalanches). These
avalanches are due to local accumulation of energy, leading to an
increasing gradient. Once the gradient exceeds an appropriate
threshold, a burst of activity, which expels the accumulated
energy, ensues. This process can be renewed, much like a domino
effect, leading to a large transport event. Specific conditions of
the thermal convection in the container make these states
metastable and produce random reversals.

\section{Multiscaling properties of the mean wind and SOC}
\label{2} The probability density function (PDF) of the life-times
of the metastable states, $\tau$, observed in the laboratory data
on the wind, exhibit the characteristic power-law
$$
p(\tau) \sim \tau^{-1},  \eqno{(1)}
$$
as shown in figure 1.

\begin{figure}
\epsfig{file=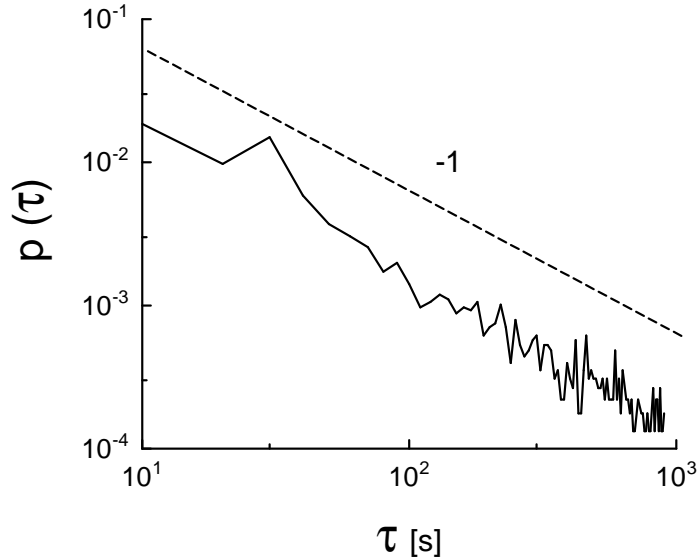,width=4.5in}
\vspace{-6cm}\caption{\footnotesize The probability density
function of the life-times of the metastable states of the wind
(in log-log scales). The dashed line indicates the power law
(1).}\vspace{1cm}

\end{figure}
Different physical mechanisms can lead to this power law. It could
be the ``exchange of stability" under turbulent noise effect
\cite{sbn}, or it could be the SOC as mentioned in the
Introduction. These mechanisms differ substantially. The
differences can be ascribed in part, almost trivially, to the lack
of long time correlations for the avalanches in the ``monoscale"
SOC models, leading to the inability of SOC to incorporate the
far-from-equilibrium characteristics of hydrodynamic systems such
as turbulent convection. This trivial distinction between the two
mechanisms disappears if we consider the two-dimensional
Bak-Tang-Wiesenfeld (BTW) model \cite{btw},\cite{dhar}, which
obeys a specific form of multiscaling for the PDF of several
avalanche measures \cite{tms},\cite{ms1},\cite{ms}. This model has
bursts with long-time correlation, and other turbulence-like
intermittent properties, if studied on the time scales of its
waves (see below). The question now is this: Can one still
distinguish between turbulence and the multiscaling SOC in this
situation? Or, are there still significant differences between the
model and the specific physical phenomenon? We shall address this
issue briefly in this paper, using as example the life-time $\tau$
of the metastable states of the mean wind.

The BTW prototype model of self-organized criticality is defined
on a square lattice \cite{dhar}. The number of ``grains" stacked
on a given site $i$ is denoted by $z_i=0,1,2,\ldots$ If $z_i<4$,
$\forall i$, the configuration is stable by construction. A random
site $k$ is selected and a grain is added to it, thus increasing
$z_k$ to $z_k+1$. If, in the process, $z_k \ge 4$, immediately
$z_k$ is reduced to $z_k-4$. This is called ``toppling". The
expelled four grains are received one each by the four nearest
neighbor site of $k$. The grains disappear if $k$ is a boundary
site. It is clear that the toppling at site $k$ may lead to
topplings at the next microscopic time step, and so on. Thus, an
avalanche made by a total number $s\geq 0$ of topplings occurs
before a new stable configuration is reached, and a new grain is
added to a random site. After many additions the system reaches a
stationary critical state in which the properties of the
avalanches are sampled.

A key notion of the BTW model is that of wave decomposition of
avalanches \cite{ikp}. The first wave is obtained as the set of
all topplings which can take place as long as the site of addition
is prevented from a possible second toppling. The second wave is
constituted by the topplings occurring after the second toppling
of the additional site takes place and before a third one is
allowed, and so on. The total number of topplings in an avalanche
is the sum of those of all its waves.  The wave size, s, has the
power-law PDF
$$
p(s)\sim s^{-1}.   \eqno{(2)}
$$

Returning now to convection, as mentioned in the Introduction, the
times taken to switch from one metastable state to the other are
very short in comparison with the interval between switchings; in
addition, the magnitude of the mean thermal wind velocity is
approximately {\it constant} during the metastable events.
Therefore, the size of the events in the experiment (in SOC terms)
is proportional to their duration; that is, $s \sim \tau$ (compare
equations (1) and (2), \cite{klgp}).

\begin{figure}
\epsfig{file=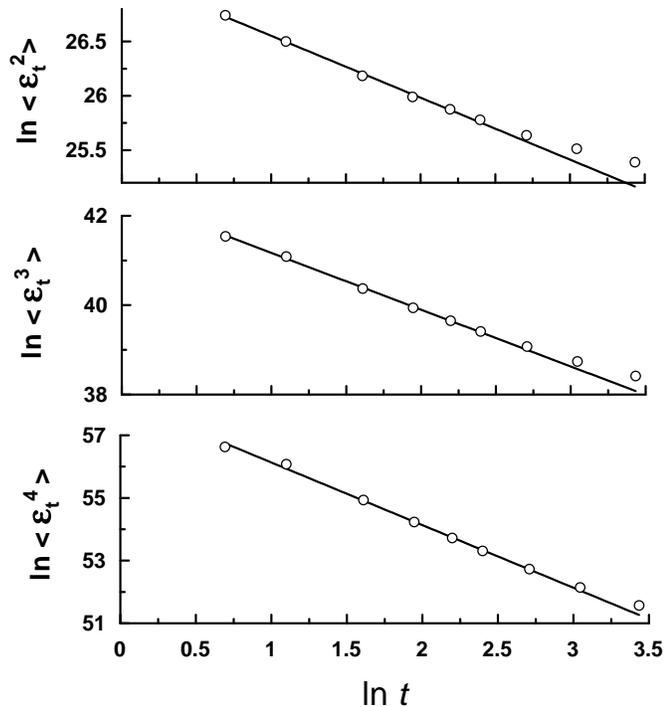,width=4.5in}
\vspace{-5cm}\caption{\footnotesize Moments of the ``local
dissipation rate" of the metastable states duration times (or, in
this case, their sizes). The straight lines (the best fit) are
drawn in order to indicate (in log-log scales) the multiscaling
given by (4).}\vspace{0.6cm}

\end{figure}

\begin{figure}
\epsfig{file=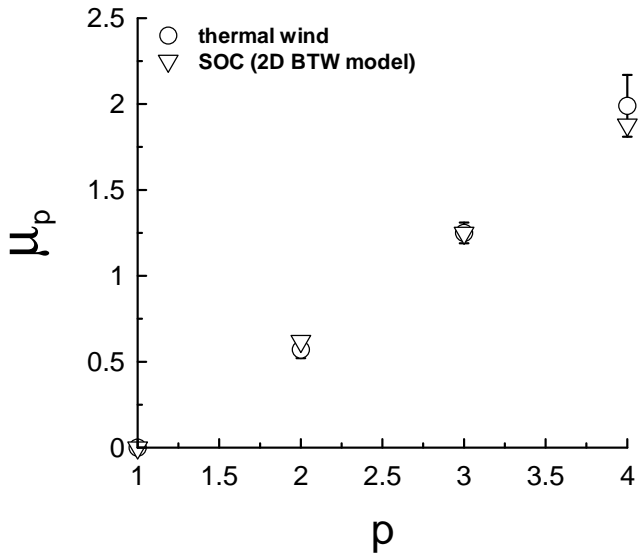,width=4.5in}
\vspace{-6cm}\caption{\footnotesize Multiscaling exponents from
(4) for the mean wind (circles) and for the two-dimensional BTW
prototype model of SOC (triangles).}\vspace{1cm}

\end{figure}

With this identification, one can take the analogy a step further.
A characteristic similar to local dissipation rate in turbulence,
$$
\epsilon_t=\sum_{k=1}^t (s_{k+1}-s_k)^2 /t,  \eqno{(3)}
$$
was introduced in the Ref. \cite{ms} in order to characterize
multiscaling properties of SOC. The multiscaling (if it exists)
has the form
$$
\frac{\langle \epsilon_t^p \rangle}{\langle \epsilon_t \rangle^p}
\sim t^{-\mu_p}.      \eqno{(4)}
$$
Such multiscaling was observed in Ref. \cite{ms} for the
two-dimensional BTW prototype model of the SOC. Since $s \sim
\tau$ in our case (see above and \cite{klgp}), we calculated the
analogous ``local dissipation rate" for the duration $\tau$
obtained in our experiment. The multiscaling and corresponding
scaling exponents $\mu_p$, given by (4), are shown in figures 2
and 3, respectively. Circles in figure 3 correspond to the
exponents computed for our data whereas the triangles correspond
to computations performed in \cite{ms} for the two-dimensional BTW
model of SOC.

\section{Concluding remarks}
\label{3}

The fluid flow here has many features such as jets, boundary
layers (both attached and detached), plumes, and so forth. These
are hydrodynamical entities which, in the end, may have to be
explained through their hydrodynamic origin. Amidst this
complexity, however, the comparison of the multiscaling exponents
shown in figure 3 suggests that the multiscale SOC is a possible
model for the observed reversal of the thermal mean wind. We might
ask: Is the mechanism in the flow indeed the same as that of
two-dimensional BTW, or does the agreement of multiscaling
exponents $\mu_p$ shown in figure 3 characterize some class of the
multiscale SOC universality? In some sense, the first question
looks for more detailed dynamical understanding, while the latter
looks for a statistical analogy. We cannot answer the first
question with any certainty yet, but the answer to the second
question seems to be in the affirmative.

\noindent{\bf Acknowledgment}

For sometime soon after SOC was invented, Per Bak was actively
proposing SOC as the leading mechanism to explain turbulence. By
the time he wrote his book \cite{bak}, however, Per seemed to have
withdrawn that emphasis. One imagines that this may have resulted
at least partly from the cautionary notes he constantly received
from the likes of one of us (KRS). It is therefore especially
fitting to point out a quantitative, albeit small, feature of
turbulence that bears similarity to SOC, even if the implications
of this observation are not yet fully clear. We appreciate the
efforts of the organizers in holding the meeting and arranging
this publication honoring the ebullient character of Per Bak. We
are pleased to be a part of it, and are sad at the same time that
his life was terminated prematurely.

\end{document}